# Subcarrier-Interlaced FDD for Faster-than-TDD Channel Tracking in Massive MIMO Systems

Maximilian Arnold, Xiaojie Wang, and Stephan ten Brink
Institute of Telecommunications, Pfaffenwaldring 47, University of Stuttgart, 70569 Stuttgart, Germany
Email: {arnold,xwang,tenbrink}@inue.uni-stuttgart.de

*Abstract*—Canonical Massive MIMO uses time division duplex (TDD) to exploit channel reciprocity within the coherence time, avoiding feedback of channel state information (CSI), as is required for precoding at the base station. We extend the idea of exploiting reciprocity to the coherence bandwidth, allocating subcarriers of a multicarrier (OFDM)-based system in an interlaced fashion to up- and downlink, respectively, referred to as (OFDM-)subcarrier interlaced FDD (IFDD). Exploiting this "two-dimensional" channel reciprocity within both time and frequency coherence does not require any CSI feedback and, even more so, offers faster-than-TDD channel tracking. To address imperfections of actual hardware, we conducted measurements, verifying the practical viability of IFDD. As it turns out, the scheme incurs similar transmit/receive isolation issues (self-interference) as well-known from the full-duplex (FDup) literature. As will be shown, such hardware challenges like self-interference or frequency offsets are not critical for massive MIMO operation, but can be neglected as the number of antennas grows large. Finally, we quantify how IFDD allows to track the channel variations much faster than TDD over a wide range of commonly used pilot symbol rates.

## I. INTRODUCTION

Massive MIMO offers unprecedented spectral efficiency with scalable transceiver complexity, and is one of the most promising technologies to fulfill the manifold requirements of future wireless systems, e.g., peak and edge data rate, massive connectivity support as well as ultra-reliable and low latency communications [1], [2], [3]. Based on precoding techniques at a basestation (BS) with a large number of antennas, the acquisition of channel state information (CSI) is carried out by transmitting uplink pilots. Therefore, massive MIMO networks favor time division duplex (TDD), as to easily exploit channel reciprocity for CSI acquisition and downlink precoding [4]. In contrast, frequency division duplex (FDD) is not a good match to massive MIMO systems, as it requires downlink channel state information to be explicitly fed back back through the uplink, i.e., from the terminals to the basestations, incurring a large overhead [5].

Despite the advantages of combining massive MIMO with TDD (referred to as "canonical massive MIMO" in the following), industry and standardization bodies have been considering FDD operation [6] as an attractive topic for 5G communications systems, partly based on the concern that channel tracking abilities in TDD are quite limited for high mobility scenarios [7], [8]. In this paper, we study a specific variant of "FDD" operation in the massive MIMO context, using the same amount of pilot overhead as in canonical massive MIMO systems. This approach, which also relies on channel reciprocity, applies an *interlaced* uplink (UL) and downlink (DL) OFDM-subcarrier allocation, also referred to as "zipped" OFDM [9]. The proposed subcarrier-interlaced FDD (IFDD) requires no explicit CSI feedback from the UE; the channel estimation is performed at the BS exploiting reciprocity, thus preserving all the advantages of canonical massive MIMO. As will be shown later, the IFDD scheme incurs similar TX/RX isolation issues as well known from research on full-duplex (FDup) systems [10]. We note that [9] focused on the investigation of channel estimation in the presence of frequency offsets with a very few number of antennas. We further address critical hardware limitations when implementing IFDD systems such as FDup interference posed by ADC saturation along with frequency offset, and benchmark those to commonly used LTE transceiver specifications. Moreover, we show that the susceptibility of IFDD to FDup interference can be mitigated by increasing the number of antennas, making "zipped" OFDM a good match for massive MIMO systems.

Obviously, the acquisition of CSI plays a crucial role in all wireless communications systems, even more so in wireless systems that rely on channel reciprocity, like massive MIMO. It is commonly assumed in the literature, that the channel remains constant within the coherence time, and the entire process of uplink data/pilot transmission, channel estimation and subsequent downlink precoding is completed within this period. However, the coherence time of wireless channels highly depends on the velocity of the mobile users. Therefore, a unified TDD or FDD frame structure needs to support a variety of users with very diverse speed profiles. The performance degradation of canonical massive MIMO systems was studied in [7], [8]; as it turns out, most losses are due to its slow channel tracking capability. Increasing the pilot rate can reduce the precoding mismatch due to the time-varying (fading) channel, which, however, sacrifices data rate and requires costly circuits supporting very fast TX/RX switching turnaround time. The IFDD solution proposed in this paper is an attractive alternative to cope with the mobility "burden" of canonical massive MIMO. Besides IFDD implementation challenges and mitigation in massive MIMO systems, this paper further addresses channel tracking capabilities of the canonical TDD-based versus the proposed IFDD-based massive MIMO system. The results show that IFDD can track the channel variations significantly faster than TDD-based massive MIMO at low to moderate pilot rates.

## II. PREREQUISITES OF INTERLACED FDD

### A. Principle of interlaced FDD

Consider an OFDM system, dividing an available bandwidth $B$ into $N_{\text{sub}}$ subchannels so that each subcarrier experiences a virtually flat fading channel. Furthermore, the BS shall be equipped with $M$ antennas while the UE has only one antenna, as commonly assumed in a canonical massive MIMO system.

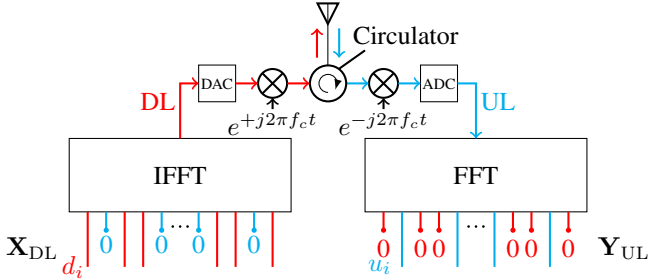

Figure 1. IFDD multicarrier transceiver structure at a carrier frequency $f_c$

The key idea of IFDD is to exploit channel reciprocity in frequency direction in addition to the time direction, consequently enabling a special type of FDD operation where the uplink and downlink share the total bandwidth with *interlaced* subcarriers as illustrated in Fig. 1; the subcarrier indices are divided into two disjoint sets $\mathcal{U}$ (uplink subcarriers) and $\mathcal{D}$ (downlink), respectively. Note that for channel reciprocity the same antenna for uplink and downlink needs to be used. The BS and UE have $M$ and one such transceivers, respectively. In the UL, pilots and data are transmitted using the uplink subcarriers $u_i$. In the DL, coherently precoded data is transmitted over the downlink subcarriers $d_i$ using the estimated uplink channel coefficients at each antenna. The downlink transmit and uplink receive RF chain can be decoupled with a circulator as shown in Fig. 1. By orthogonality of OFDM subcarriers under ideal conditions, the UL and DL signals can be perfectly separated. Hardware imperfections compromising orthogonality will be discussed later in Section III. It is noteworthy that the IFDD scheme requires no feedback from the UE to the BS, and thus, the total signaling overhead is exactly the same as in TDD, i.e., solely depends on the total number of users but not on the number of antennas $M$, while the CSI acquisition overhead of other FDD operations more or less scales with $M$.

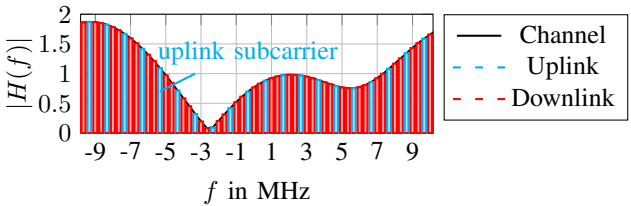

Figure 2. Exploiting channel reciprocity in IFDD across adjacent subcarriers

Let $u_i \in \mathcal{U}$ and $d_i = u_i - 1$, $d_{i+1} = u_i + 1 \in \mathcal{D}$ denote an arbitrary uplink subcarrier, and two adjacent downlink subcarriers, respectively. The estimate of the channel transfer function (CTF) of the $m$-th OFDM symbol at the $k$-th antenna $\hat{H}_{m,k}(u_i)$ constitutes a good precoding element for the downlink data transmission using the subcarriers $d_i$ and $d_{i+1}$, starting from the subsequent OFDM symbol $m+1$. It is assumed that the uplink and downlink subcarriers are within the coherence bandwidth of the channel (which will be further explained in Sec. II-B). The channel coefficients are estimated at the BS through uplink pilots. For simplicity of notation, and without loss of generality, a single user is considered. After conventional OFDM receiver processing, the channel vector containing the channel coefficients of all antennas and at the $u_i$-th subcarrier is expressed by

$$\hat{\mathbf{h}}_m(u_i) = \frac{\mathbf{Y}_{\text{UL}}(u_i) + \mathbf{X}_{\text{DL}}(u_i)}{S_{\text{p}}(u_i)}, \quad (1)$$

where $\mathbf{Y}_{\text{UL}}(u_i)$, $\mathbf{X}_{\text{DL}}(u_i)$ and $S_{\text{p}}(u_i)$ denote the received uplink symbol, transmitted downlink symbol vector at all antennas, and the known pilot symbol at the $u_i$-th subcarrier, respectively. Ideally, $\mathbf{X}_{\text{DL}}(u_i) = 0$, as per orthogonality of the OFDM subcarriers. After CSI acquisition, the DL data can be precoded using, e.g., maximum ratio (MR) precoding, according to

$$\mathbf{X}_{m'}^{\text{DL}}(d_j) = \beta \hat{\mathbf{h}}_m^H(u_i) X_{m'}(d_j), \quad (2)$$

where $\beta$ scales the transmit power, $m' \in \{m+1, \cdots, m+K_c-1\}$ denotes the symbol index within the channel coherence time ($K_c$ is the coherence time in terms of OFDM symbols) and $d_j \in \{d_i, d_{i+1}\}$ is an arbitrary DL subcarrier index within the coherence bandwidth of the $u_i$-th UL subcarrier.

### B. Essential requirements for interlaced FDD

IFDD requires that the UL and DL subcarriers reside within the coherence bandwidth of the channel, which can be achieved by increasing the number of subcarriers $N_{\text{sub}}$, i.e., decreasing the subcarrier spacing. We quantify the channel correlation between uplink and adjacent downlink subcarriers as

$$\delta_h(\Delta f, \Delta t) = \frac{\left\|\mathbf{h}_m(u_i)\mathbf{h}_{m+\Delta t}^H(d_j)\right\|}{\sqrt{\left\|\mathbf{h}_m(u_i)\right\|^2 \left\|\mathbf{h}_{m+\Delta t}(d_j)\right\|^2}}, \quad (3)$$

where $\Delta f = d_j - u_i$ and $\Delta t$ are both normalized to the subcarrier spacing and the OFDM-symbol duration, respectively. Note that the two channel vectors differ in both time and frequency domain. This channel correlation measure corresponds to the precoding mismatch of maximum-ratio precoding (MR) which makes this correlation model a suitable measure for quantifying the mismatch between up- and downlink channels.

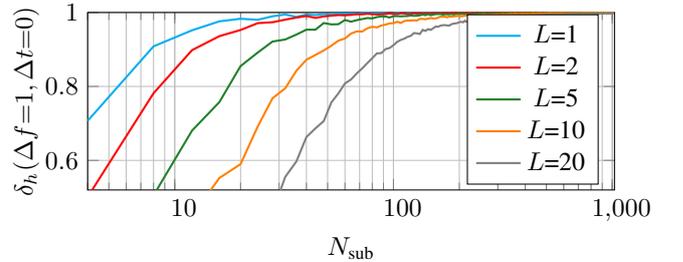

Figure 3. Channel correlation $\delta_h$ versus number of subcarriers $N_{\text{sub}}$ at a fixed bandwidth for different channel orders $L$

Fig. 3 shows the channel correlation over the number of subcarriers with different channel order $L$ (corresponding to the length of the respective tapped-delay line model), averaged over 1000 channel realizations. The duration of each tap is $1/T_s = 1/(T_{\text{OFDM}} N_{\text{sub}})$. Each tap of the channel is normalized to the power $\frac{1}{L+1}$ so that there is no (i.e., 0dB) channel gain. As can be seen, with increasing number of channel taps $L+1$, the frequency selectivity increases, and thus, more subcarriers per fixed overall bandwidth are required to maintain high correlation among adjacent subcarriers, which, in turn, allows to exploit channel reciprocity for precoding in IFDD operation. While $N_{\text{sub}}$ can be designed arbitrarily large to achieve large coherence bandwidth in terms of number of subcarriers, its upper bound is imposed by the channel coherence time $T_c$. Suppose that the minimum number of subcarriers within the channel coherence time is 3, since one uplink subcarrier is used to estimate the two adjacent downlink channels (which is the minimal coherence bandwidth required to use IFDD), i.e., $B_c \geq 3 \frac{B}{N_{\text{sub}}}$, and at least two symbols shall be transmitted within $T_c$, i.e., $T_c \geq 2 \frac{N_{\text{sub}}}{B}$. Thus, IFDD works fine as long as

$$\frac{T_c}{2} \geq \frac{N_{\text{sub}}}{B} \geq \frac{3}{B_c}, \quad (4)$$

where $B$ denotes the signal bandwidth. The operation range of TDD, which can be easily calculated as $\frac{T_c}{2} \geq \frac{N_{\text{sub}}}{B} \geq \frac{1}{B_c}$, is wider than that of IFDD as merely the coherence time is required; the channel in frequency direction only needs to be "flat" across one subcarrier. We note that the subcarrier spacing in LTE is 15kHz, and the typical wireless channel coherence bandwidth is about 50-500kHz, with 2-200ms coherence time, respectively, thus fulfilling above prerequisite (4) of IFDD operations without the need of increasing $N_{\text{sub}}$.

## III. A PROOF-OF-CONCEPT IMPLEMENTATION OF IFDD

Theoretically, both UL and DL are mutually orthogonal in IFDD, although in practical systems, hardware impairments can compromise orthogonality, inducing severe intercarrier interference (ICI). Fig. 4 shows the effect of typical hardware impairments and a local scatterer in IFDD operation.

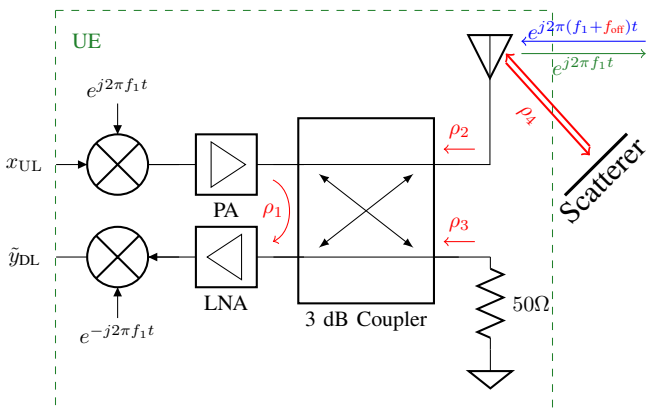

Figure 4. Dominant hardware challenges for IFDD (TX/RX cross-talk), also referred to as self-interference, FDup, quantified as isolations $\rho_{1,..,4}$ at the user equipment (UE)

To exploit channel reciprocity, it is required to use the *same* antenna for up- and downlink in massive MIMO systems. Therefore, a circulator or 3dB-coupler is needed to isolate the transmit and the receive path. Since neither coupler nor circulator are able to separate both paths perfectly, crosstalk with an isolation $\rho_1$ from the transmit path to the receive path remains. In addition, reflections of the transmitted signal occur because of mismatched antenna and load over the used bandwidth at both paths. These reflections are modeled by the backward leakage $\rho_2$ at the antenna and $\rho_3$ at the load, respectively. Moreover, a carrier frequency offset of $f_{\text{off}}$ is present due to spatially separated local oscillators of BS and UE, respectively. Also, any local scatterer around the UE causes additional reflections of the transmitted signal resulting in yet another source of crosstalk between UL and DL, referred to as $\rho_4$. For the BS, the condition of "no local scatterers in the vicinity" is often satisfied, and the analysis at the BS is similar but simpler compared to the UE. Therefore, we focus on the more critical scenario at the UE. In the following, we discuss the influences of these impairments in detail, not only by theoretical considerations but also by measurements in a SISO system testbed, which is sufficient to show the main challenges and verify the viability of the scheme.

### A. Influence of frequency offsets

As depicted in Fig. 4, the UE receives and decodes the DL signal from the BS while simultaneously transmitting its UL signal. Because of the interlaced subcarrier allocation for UL and DL, IFDD is quite sensitive to carrier frequency offset (CFO), which causes interference similar to a full-duplex operation. In the following we consider the impact of frequency misalignment on the demodulation of DL signals at the UE. Suppose that the DL signal arrives at the UE with the propagation delay of $\tau_{\text{ch}}$, the delays and attenuations of reflections at the 3dB coupler, antenna, load and the local scatterer are $\tau_{r,1}, \tau_{r,2}, \tau_{r,3} \, \tau_{r,4}$ and $\rho_1, \rho_2, \rho_3, \rho_4$, respectively. The superimposed signal after down-conversion is given by

$$\tilde{y}_{\text{DL}}(t) = e^{j2\pi f_{\text{off}} t} y_{\text{DL}}(t - \tau_{\text{ch}}) + \sum_{i=1}^{4} \rho_i x_{\text{UL}}(t - \tau_{r,i}), \quad (5)$$

where $y_{\text{DL}}(\cdot)$ represents the DL after passing through the channel, and $x_{\text{UL}}(\cdot)$ is the UL signal to be transmitted, right before the channel. Assume that the inserted cyclic prefix (CP) $t_{\text{cp}}$ is larger than the maximum delay time among the propagation and all reflections, which writes as

$$t_{\text{cp}} \geq \max_{\forall i \in [1,4]} \{\tau_{\text{ch}}, \tau_{r,i}\}. \quad (6)$$

It can be expected that the maximum delays of those reflections are much smaller than the propagation delay of the received signals, i.e., $\max_{\forall i \in [1,4]} \{\tau_{\text{ch}}, \tau_{r,i}\} = \tau_{\text{ch}}$. Therefore, IFDD requires no additional CP overhead to preserve orthogonality. In fact, the amount of CP overhead in IFDD can be reduced because the OFDM symbol duration is, by design, larger than that of TDD. After some mathematical manipulations it can be shown that the signal at an arbitrary subcarrier $\ell$ writes as

$$Y(\ell) = \underbrace{\alpha_{\text{rx}} X_{\text{DL}}(\ell) H(\ell) G(f_{\text{off}}) e^{j2\pi\ell \frac{t_{\text{cp}}-\tau_{\text{ch}}}{T_s}}}_{\text{desired signal}}$$

$$+ \underbrace{\alpha_{\text{rx}} \sum_{\substack{n=0 \\ n\neq\ell}}^{N_{\text{sub}}-1} X_{\text{DL}}(n) H(n) G((n-\ell)f_{\text{sub}} + f_{\text{off}}) e^{j2\pi n \frac{t_{\text{cp}}-\tau_{\text{ch}}}{T_s}}}_{\text{inter carrier interference}}$$

$$+ \underbrace{\alpha_{\text{tx}} \sum_{i=1}^{4} \rho_i X_{\text{UL}}(\ell) G(0) e^{j2\pi\ell \frac{t_{\text{cp}}-\tau_{r,i}}{T_s}}}_{\text{full duplex interference = 0}}, \quad (7)$$

where $G(\cdot)$ is the pulse shaping of each OFDM subcarrier, $f_{\text{sub}} = \frac{N}{T_s}$ denotes the subcarrier spacing and $\alpha_{\text{tx}} = \sqrt{P_{\text{tx}}/N_{\text{sub}}}$, $\alpha_{\text{rx}} = \sqrt{P_{\text{rx}}/N_{\text{sub}}}$ correspond to the transmitted and received signal power, respectively. A direct consequence of (7) is that the "full-duplex" interference vanishes by orthogonality of subcarriers, provided that a sufficiently long cyclic prefix in (6) is used. Hence, the average Signal-to-Interference-Ratio (SIR) can be computed as

$$\overline{\text{SIR}}(\ell) = \frac{|G(f_{\text{off}})|^2}{\sum_{\substack{n=0 \\ n\neq\ell}}^{N_{\text{sub}}-1} |G((n-\ell)f_{\text{sub}} + f_{\text{off}})|^2} \quad (8)$$

with normalized symbol power $\mathrm{E}\left[|X_{\text{DL}}(i)|^2\right] = 1$ and channel gain $\mathrm{E}\left[|H(i)|^2\right] = 1$. Furthermore, the transmitted QAM symbols at different subcarriers are assumed to be uncorrelated, i.e., $\mathrm{E}\left[|X_{DL}(i) \cdot X_{DL}^*(j)|\right] = 0, \forall i \neq j$.

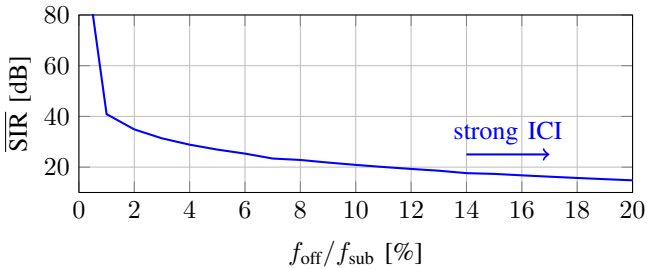

Figure 5. Average SIRs versus normalized CFO for an OFDM system with 1024 subcarriers at the second (arbitrary) subcarrier

Fig. 5 shows the achievable SIR over a range of normalized CFOs (Inter carrier interference ICI) for an OFDM system with $N_{\text{sub}} = 1024$. Obviously, for a frequency offset of 1-5% with respect to the subcarrier spacing, IFDD can achieve a reasonable high SNR.

### B. The full-duplex ADC challenge

Considering that orthogonality between UL and DL is maintained thanks to the CP, the aforementioned reflections of the transmitted signal pose a great challenge to the analog-digital converter (ADC) at the RX frontend: the power of the reflected "self-interference" is orders of magnitudes larger (i.e., several 10s of dB larger) than the power of the actually desired received signal. To address this problem, we consider the signal to quantization noise ratio (SQNR). Assuming an equally distributed quantization error, a triangular-shaped amplitude distribution of the signal $x_{\text{UL/DL}}$, and using the full dynamic range of the ADC, the achievable SQNR can be calculated as

$$\gamma_Q = \frac{1.5N^2}{1 + \rho \frac{P_{\text{tx}}}{P_{\text{rx}}}} \quad (9)$$

with $N$ being the total number levels of the ADC and $\rho = \sum_i \rho_i$ being the worst-case estimation of the leakage, where each impairment (see Fig. 4) adds up constructively. Note that $\rho$ is usually in the range of -10 to -36 dB depending on hardware parameters and local scatterers (see reflection coefficient in [11]). The difference in power by orders of magnitudes between transmitted and received signals substantially limits the achievable SQNR. For instance, the SQNR of the received signal can be as small as $-40$dB in a SISO-case with $N = 256$, $\rho = -10$ dB and $\frac{P_{\text{tx}}}{P_{\text{rx}}} = 100$ dB, rendering IFDD useless for SISO. However, as will be shown next, the SQNR issue can be mitigated by the the larger number of antennas in massive MIMO, making IFDD and massive MIMO a very good match. With MR precoding, as an example, the required transmit power at each basestation antenna ([4], p. 116) is proportional to $1/M^{\varepsilon_1}$, while the uplink power is proportional to $1/M^{\varepsilon_2}$ for achieving a non-zero asymptotical spectral efficiency, given that $\varepsilon_1 + \varepsilon_2 < 2$. With this transmit power strategic in massive MIMO, the achievable SQNRs can be derived as shown in Tab. I. We note two effects of increasing the number of basestation antennas, weakening the full-duplex interference: 1.) In the downlink the precoding effect (array gain) of the large array increases the received power which drastically decreases the difference between the transmit and received power at the UE; 2.) the transmit power per antenna can be reduced by increasing the number of antennas which reduces the self interference while still achieving a spectral efficiency gain. We infer that, to reduce the full-duplex problem by a factor close to $1/M$, we need to have $\varepsilon_1 - \varepsilon_2 = 1$. To achieve non-zero asymptotical spectral efficiency and simultaneously decline the full-duplex problem, we obtain that $\varepsilon_1 = 3/2$ and $\varepsilon_2 = 1/2$. Therefore, the SQNR, in all cases, improves by $1/M$ with increasing the number of basestation antennas, while the system still maintains its spectral efficiency gains.

Table I
ACHIEVABLE SQNR AT DIFFERENT TRANSMISSION STAGES IN IFDD

|  | Uplink Pilot/Data | Downlink Data |
|---|---|---|
| $\gamma_Q=$ | $\dfrac{1.5N^2}{1+\dfrac{\rho}{M^{\epsilon_1-\epsilon_2}}\cdot\dfrac{P_{\text{tx}}}{P_{\text{rx}}}}$ | $\dfrac{1.5\cdot N^2}{1+\dfrac{\rho}{M^{2-\epsilon_1+\epsilon_2}}\cdot\dfrac{P_{\text{tx}}}{P_{\text{rx}}}}$ |

Next we consider an LTE-transceiver example (c.f. [12]): Consider an LTE device with an 8bit-ADC in a rich scattering (RS) environment ($\rho = -10$dB) and an environment without rich scattering in the vicinity ($\rho = -30$dB), respectively. Furthermore, an automatic gain control for adjusting the input attenuation (or amplification) is needed for TDD and IFDD. Fig. 6 shows the achievable SQNR versus the number of basestation antennas.

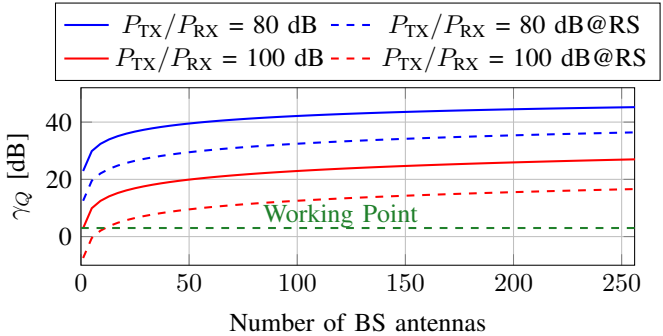

Figure 6. Achievable SQNR for the uplink pilots versus the number of basestation antennas

As depicted in Fig. 6, the SQNR is in a reasonable range even if local scatterers are nearby a UE. Moreover, the "Working Point" is an operating SNR of an LTE system with a reasonable link quality, which is easily achieved by using 16 BS antennas. Further increasing the BS antenna number leads to negligible FDup interference compared to noise.

### C. Measurement setup for TDD and IFDD

To demonstrate the viability of IFDD with real hardware impairments, we devised a simple SISO-measurement setup, as depicted in Fig. 7.

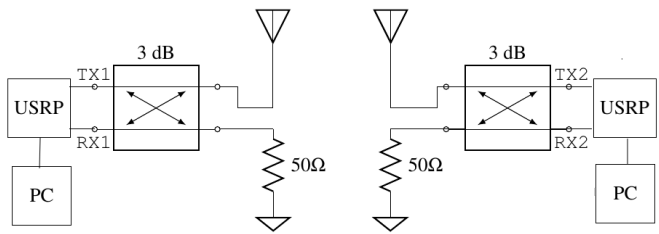

Figure 7. Measurement setup for evaluating channel reciprocity under real hardware impairments using two software defined radios (SDRs, USRPs)

First we consider channel reciprocity measurements of a TDD system, as the IFDD system also relies on channel reciprocity within the coherence time. For the TDD measurement, the two universal software defined radios (USRPs: B200 and B210) [13] transmit, each, two OFDM symbols with 1024 subcarriers in an alternating fashion as prescribed by the TDD protocol. A full cycle of uplink, downlink and transient time is 1ms. Next, the channel is estimated over the OFDM symbols with 20MHz bandwidth (the SNR was ensured to always be above 20dB). The transmitter and receiver are synchronized via an GPSDO (GPS disciplined oscillator). The measurement was conducted in a typical office environment (c.f. [14]) on a desk (for the NLoS scenario a metal plate was inserted between the USRPs). For convenience, the distance between the USRPs was fixed for TDD at 1.5m.

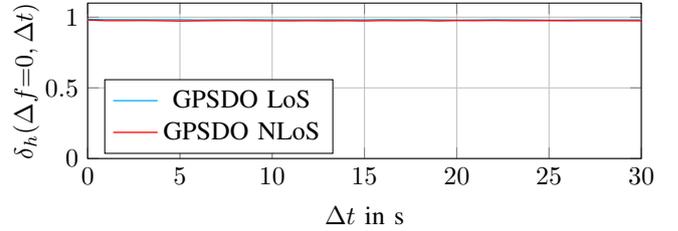

Figure 8. Measured: Channel reciprocity for TDD expressed in terms of $\delta_h$ versus time averaging over all subcarriers

Fig. 8 quantifies the "channel reciprocity" in terms of the correlation coefficient as defined in (3) of uplink and downlink in the LoS/NLoS scenario for TDD. The channel correlation coefficient is measured between uplink and downlink channel over time. It can be observed that the TDD assumption (reciprocity maintained over time) holds very well. Thus, after these initial sanity checks, the hardware is ready for testing the more involved case of IFDD.

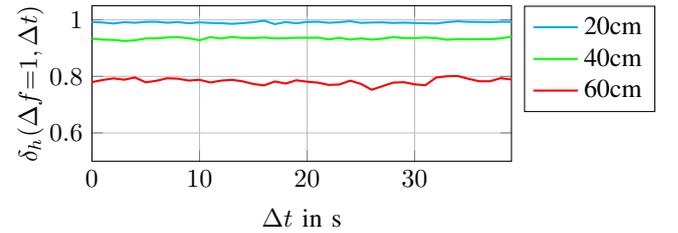

Figure 9. Measured: Channel reciprocity expressed in terms of $\delta_h$ for IFDD of adjacent subcarriers versus time; at a distance of 20,..,60cm between TX and RX

The measurement in Fig. 9 for the IFDD experiment depicts the reduction of channel correlation caused by the growing power difference between transmitting and receiving (transmitter and receiver are separated with the distance of 20, 40, 60 cm, respectively). This is due to the fact that with higher distance the transmit power of the USRP needs to be increased to maintain a constant receive power. Therefore the self interference increases and the ADC saturates. The channel correlation coefficients are measured between adjacent uplink and downlink subcarriers over time, which quantify the reciprocity in time and frequency direction expressed over $\delta_h$. Moreover, it is shown that IFDD is practically possible, since, within 20cm, the channel correlation between the actual downlink and the estimated uplink is close to 1, i.e., $\delta_h \approx 1$. At 60cm IFDD suffers heavily from the full-duplex problem in time domain, which, however, can be mitigated through increasing the number of antennas at the basestation (see Fig. 6). This measurements lead to the conclusion that IFDD with massive MIMO could be a practical solution simultaneously enabling FDD while suppressing FDup interference.

## IV. FASTER-THAN-TDD CHANNEL TRACKING

Fast channel tracking is desired due to the time-varying nature of wireless channels, particularly in systems such as massive MIMO that exploit channel reciprocity to precode towards mobile users. Mismatch between the acquired CSI at transmitter and the actual channel is inevitable due to, e.g., the users' motion. While massive MIMO offers excellent immunity against CSI corrupted by uncorrelated noise, the precoding mismatch due to outdated CSI may obliterate all of its advantages. In the following, we present simulation results considering the CSI mismatch caused by users' motion and compare the channel tracking capabilities of TDD and IFDD.

### A. Simulation settings

Tab. II shows some LTE-like system parameters (c.f. [12]) as used in the subsequent numerical performance evaluation.

Table II
SIMULATION PARAMETERS

| Simulation parameter | Value |
|---|---|
| $N_{\text{sub,TDD}}$ | 1024 |
| $N_{\text{sub,IFDD}}$ | 2048 |
| CP length | 128/256 samples |
| Carrier frequency | 2.1GHz |
| Sampling rate | 20MHz |
| TDD transient time ($\tau_{\text{ST}}$) | 20 samples, i.e., $1\mu s$ |
| Doppler spectrum | Classic (Jakes & Clark) |
| Channel order | 10 |
| Modulation scheme | QPSK |
| SNR (see working point in Fig. 6) | 3dB |
| Number of users | 1 |
| BS antennas | 128 |

The number of subcarriers of IFDD is chosen to be twice the one of the corresponding TDD system, even though the LTE system with subcarrier spacing of $15\,\text{kHz}$ is mostly overdimensioned in terms of coherence bandwidth, e.g., for this channel the 90-percent coherence bandwidth is 120kHz. In TDD, a transient time of 1µs between uplink and downlink is needed, which is practically possible as, e.g., demonstrated in [15]. Fig. 10 shows the protocol setup, for instance, with the pilot rate $p = \frac{1}{3}$ defined as the ratio between pilot and total uplink data.

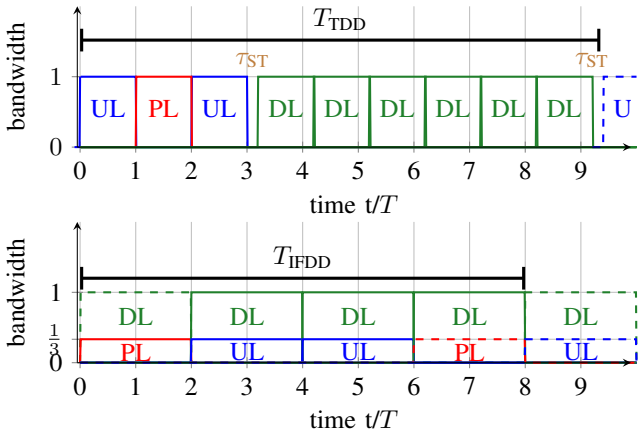

Figure 10. Simulation setup for a 33.33% pilot rate

Note that the symbol duration for TDD is $T = T_{\text{CP}} + T_{\text{symbol}}$ and for IFDD is $2T$. The downlink data rate is set to be twice the one of the uplink, modeling a realistic demand. The uplink pilots are placed in the middle of the uplink frame, since it is a good trade-off between uplink data equalization and downlink precoding. The required time for transmitting and receiving each TDD and IFDD frame can be calculated as

$$T_{\text{TDD}} = \underbrace{\frac{2T}{p}}_{\text{DL}} + \underbrace{\frac{T}{p}}_{\text{UL \&PL}} + 2\tau_{\text{ST}}$$

$$T_{\text{IFDD}} = \frac{2T}{p} + \underbrace{2T}_{\text{PL}}.$$

Obviously, the transient time in TDD ($\tau_{\text{ST}}$) always decreases the data rate. We assume rich scattering at the UE, thus the time-varying characteristic of the channel is the classic Jakes' spectrum. To evaluate the channel tracking capabilities of both systems, the achievable (bit-interleaved coded modulation, BICM [16]) mutual information of the DL data based on the estimated and mismatched CSI is compared. The impact of pilot rate $p$, number of antennas $M$ and the channel fading rate $\nu$ on the achievable rates $I$ is investigated at a desired SNR by computing the

$$I_{\text{TDD}}(p, M, \nu, \text{SNR}) = \frac{1}{B \cdot T_{\text{TDD}}} \sum_i I\left(\hat{b}_i; b_i\right)$$

$$I_{\text{IFDD}}(p, M, \nu, \text{SNR}) = \frac{1}{B \cdot T_{\text{IFDD}}} \sum_i I\left(\hat{b}_i; b_i\right);$$

where $I\left(\hat{b}_i; b_i\right)$ is the average mutual information, in bits per channel use (bpcu), between the hard decided bits and the transmitted bits, where $B$ is the total bandwidth used.

### B. Comparison of TDD and IFDD

We show in the following the channel tracking ability of both systems in various fading evironments, i.e., the user is moving at the speed of 10, 45 and 100 kmh, respectively. The number of antennas at the basestation is 128. For fair comparison, the pilot overhead or pilot percentage is kept the same for both systems. The achievable information rates in the downlink are simulated over different pilot percentages, taking into account the precoding mismatch and TDD transient time. From Fig. 11 it can be seen that the achievable rate of TDD reaches its maximum in the slow and moderate fading environment at the pilot rate of 33% and 50%, respectively, since the acquired CSIs are up to date. Further increase of the pilot rate only sacrifices data rate and increases the loss due to TDD transient switching time. In contrast, IFDD always benefits from higher pilot rates. Furthermore, IFDD outperforms TDD at all pilot rates in the slow and moderate fading cases while TDD shows its advantage with extremely large pilot rates of 50% and 100% in the fast fading case. If lower pilot rates are used in the fast fading case, IFDD still offers better performance than TDD.

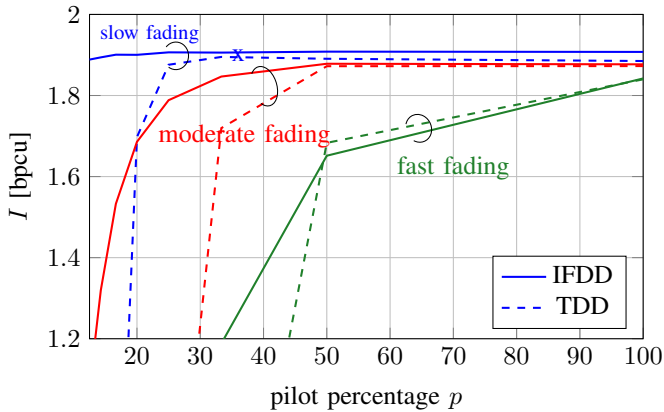

Figure 11. TDD vs IFDD in a slow, moderate and fast fading environment at various pilot rates

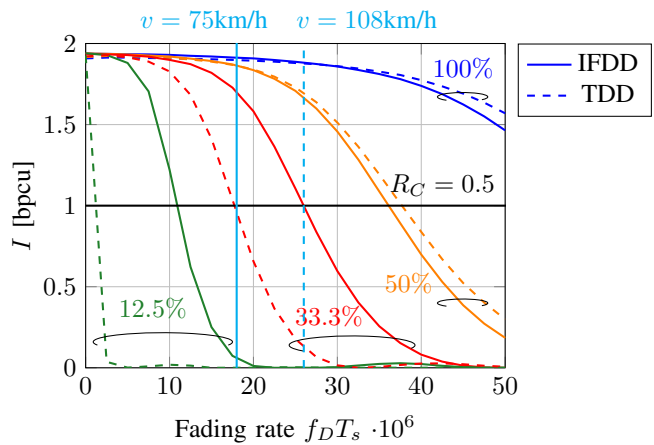

Figure 12. TDD vs IFDD at different channel fading rates and pilot rates

Fig. 12 shows the influence of the channel fading rate on TDD and IFDD at the pilot rates of 12.5, 33.3, 50 and 100%, respectively. In the fast fading regime, TDD outperforms IFDD at unrealistically high pilot rates of 100% and 50% (both only given for reference), mainly due to its shorter OFDM symbol duration compared to IFDD. At more realistic pilot rates such as 33.3% and 12.5%, IFDD is far more robust against channel variations, as enabled by the fast procedure of CSI acquisition (c.f. Fig. 10). Moreover, IFDD achieves higher rates in the slow fading regime at all pilot rates. As an example, at a given pilot rate of 1/3 and a coding rate of 1/2, TDD can support users up to 75km/h, while IFDD can improve the mobility support by 44% thus serving users at the speed up to 108km/h.

## V. Conclusion

We proposed an interlaced frequency division duplex (IFDD) scheme in a massive MIMO context, exploiting channel reciprocity in the time and frequency direction (i.e., across adjacent subcarriers). The IFDD scheme is more susceptible to hardware impairments such as frequency offset and self-interference (TX/RX isolation akin to the well-known full-duplex problem). We show that IFDD and massive MIMO are a very good match as IFDD extends the mobility support of massive MIMO, while, in turn, the large number of antennas mitigates the full-duplex problem of IFDD. Both, channel reciprocity in time and frequency direction, as exploited in IFDD, and the main implementation challenges were verified by measurements in a simple SISO link setup. Moreover, design rules for implementing IFDD in current network scenarios are provided. Finally, we compared the channel tracking abilities of IFDD and TDD taking into account the mismatch between precoder and the actual channel due to the time-varying nature of wireless channels. We observed that channel tracking of IFDD is much faster than that of TDD over a wide range of practically relevant pilot rates, making it an attractive candidate for improving the mobility support of future massive MIMO systems.